# Will the Real 4D, $N = 1$ SG Limit of Superstring/M-Theory Please Stand Up?[1]

S. James GATES, Jr.[2]   and   Hitoshi NISHINO[3]

*Department of Physics*
*University of Maryland at College Park*
*College Park, MD 20742-4111, USA*

**Abstract**

At the component-level we study the 'beta-function-favored constraint' ($\beta$FFC) formalism, suggested in 1988 as the most natural formulation for supergravity derived from more fundamental theories. We begin with the suggestion that $\beta$FFC supergravity be identified with new minimal supergravity together with an additional chiral compensator multiplet. After $U_{\rm A}(1)$-symmetry breaking, the non-propagating axion 2-form of new minimal supergravity becomes the propagating axion 2-form required by string theory. The final form of the theory is seen to *uniquely* allow four simultaneous features: (i) local supersymmetry, (ii) implementation of the Green-Schwarz mechanism, (iii) a supersymmetry-breaking order parameter chiral superfield, and (iv) a dilaton superpotential.

---

[1]This work is supported in part by NSF grant # PHY-98-02551.
[2]E-Mail: gatess@wam.umd.edu
[3]E-Mail: nishino@nscpmail.physics.umd.edu

## 1. Introduction

Much of the 'conventional wisdom' regarding superstring and heterotic string theory is based on unstated assumptions (without the benefit of supporting rigorous mathematically-based calculations) that are seldom questioned in many investigations. In such an era, we believe it is of value to look for results that defy the orthodoxy. Owing to a lack of a complete covariant heterotic string field theory (or superstring field theory) formalism, almost all of the literature on 'string-inspired phenomenology' is strewn with large numbers of such assumptions. In particular, there is a specific example to which we wish to again turn our attention in this present work. In the following, we will explore, in much greater detail than previous expositions, our proposal that the 4D, $N = 1$ supergravity limit of heterotic and superstring theories is an unconventional theory distinctly different from that used for traditional string-inspired phenomenology models. For 4D, $N = 1$ supergravity, there are two distinct off-shell formulations with the minimal number of auxiliary fields called 'old minimal' [1] and 'new minimal' supergravity [2][3][4][5], respectively. These two sets of auxiliary fields are understood as two different ways of gauge-fixing 4D, $N = 1$ conformal supergravity.

On the other hand, among infinitely many sets of constraints in 4D, $N = 1$ superspace formulations connected by super-Weyl rescalings, there is a very peculiar set of constraints called 'beta-function-favored constraints' ($\beta$FFC) [6][7][8] which drastically simplifies the beta-function computations for Green-Schwarz superstring $\sigma$-models [9]. This set of constraints was first developed [6] for the on-shell formulation of 10D, $N = 1$ supergravity, and later developed also in 4D [7][8] as an off-shell formulation for $N = 1$ supergravity. After our initial suggestion regarding the $\beta$FFC formulation, two other investigations [10] (also carefully studying aspects of the manifest realization of target space supersymmetry within string theory) found indications that $\beta$FFC supergravity is indeed the 4D, $N = 1$ supergravity limit of heterotic and superstrings. However, the explicit connection between these three off-shell formulations, *i.e.,* old minimal, new minimal and $\beta$FFC formulations has never been elucidated at the level of component fields.

In this Letter, we show at the component level the direct link between these supergravity formulations, by studying the $\beta$FFC formulation [7].[4] We first recognize that the $\beta$FFC formulation in [7] can be identified with new minimal supergravity [2] with chiral or linear multiplets, up to appropriate super-Weyl rescalings, by studying the relevant superspace commutator algebra. To study $\beta$FFC supergravity more closely, we next consider the kinetic Lagrangian for a chiral multiplet, with the 'wrong' sign for its kinetic term, coupled to new minimal supergravity [2] accompanied by the usual Brans-Dicke term. We next perform a field-redefinition on the auxiliary axial gauge field $A_\mu$ for the axial $U_A(1)$-symmetry, by eliminating a bilinear mixing of the fields. We then show that this step yields a new kinetic term for the antisymmetric tensor $a_{\mu\nu}$ which was originally just an auxiliary field (with no propagating physical degrees of freedom) in new minimal supergravity [2]. As is also desired, the kinetic term for the scalar field $A$ in the chiral multiplet acquires the right sign for its kinetic term, after a Weyl-rescaling to recast the Brans-Dicke term into the usual scalar curvature term. Similarly, the kinetic term for the pseudoscalar field $B$ disappears, whose degree of freedom is now transferred into that of the new propagating axion field $a_{\mu\nu}$ required by string theory and conserving the on-shell physical degrees of freedom. This is to our knowledge a unique example of a Goldstone-like mechanism acting between fields of the same spin and between propagating and auxiliary fields. The auxiliary 2-form 'eats' the pseudo-scalar in the compensating multiplet and itself becomes physical. We also see that this mechanism works only for a special range of the $U_A(1)$ charge of the chiral compensator multiplet. Our result is also consistent with the conclusion by Siegel [13] which dictates that the BRST structure of superstring theory implies the presence of a chiral compensating superfield in $N=1$ supergravity theory in 4D. This result forbids the interpretation of [14] from being correct.

---

[4]Even though there have been works giving [11][12] relationships among these theories at the component level, the direct link between them (in particular combined with the superstring *via* $\beta$FFC) has never been elucidated as we will show in the present paper.

## 2. A Formulation of New Minimal Supergravity

We first study the essential structure of the $\beta$FFC formulation of 4D, $N=1$ supergravity [7][8] in superspace. The commutator algebra for purely supergravity sector is generated by the supercovariant derivative

$$\nabla_A \equiv E_A{}^M \partial_M + \tfrac{1}{2}\phi_{Ab}{}^c \mathcal{M}_c{}^b - iA_A \mathcal{Y} , \tag{2.1}$$

where $\mathcal{M}_a{}^b$ is the Lorentz generator and $\mathcal{Y}$ is the $U_A(1)$-symmetry generator in the system gauged by the potential superfield $A_A$ [7][8]. The corresponding commutator algebra is given by [8][7][5]

$$[\nabla_\alpha, \nabla_\beta\} = 0 , \tag{2.2a}$$

$$[\nabla_\alpha, \overline{\nabla}_{\dot\alpha}\} = i\nabla_a + H_{\beta\dot\alpha} \mathcal{M}_\alpha{}^\beta - H_{\alpha\dot\beta}\overline{\mathcal{M}}_{\dot\alpha}{}^{\dot\beta} + H_a \mathcal{Y} , \tag{2.2b}$$

$$[\nabla_\alpha, \nabla_b\} = i(\nabla_\beta H_{\gamma\dot\beta})( \mathcal{M}_\alpha{}^\gamma + \delta_\alpha{}^\gamma \mathcal{Y})$$
$$+ i[\, C_{\alpha\beta} \overline{W}_{\dot\beta\dot\gamma}{}^{\dot\delta} - \tfrac{1}{3}\delta_{\dot\beta}{}^{\dot\delta}(2\nabla_\alpha H_{\beta\dot\gamma} + \nabla_\beta H_{\alpha\dot\gamma})\,]\overline{\mathcal{M}}_{\dot\delta}{}^{\dot\gamma} , \tag{2.2c}$$

$$[\nabla_a, \nabla_b\} = \Big[\, \{\, \tfrac{1}{2}C_{\alpha\beta}[\, i H^\gamma{}_{(\dot\alpha|}\nabla_{\gamma|\dot\beta)} - \tfrac{1}{2}([\nabla^\gamma, \overline{\nabla}_{(\dot\alpha|}]H_{\gamma|\dot\beta)} + i(\nabla^\gamma{}_{(\dot\alpha|}H_{\gamma|\dot\beta)}))\mathcal{Y}\,]$$
$$+ [\, C_{\dot\alpha\dot\beta}(W_{\alpha\beta}{}^\gamma - \tfrac{1}{6}(\overline{\nabla}^{\dot\gamma}H_{(\alpha|\dot\gamma)}\delta_{|\beta)}{}^\gamma) - \tfrac{1}{2}C_{\alpha\beta}(\overline{\nabla}_{(\dot\alpha}H^\gamma{}_{\dot\beta)})\,]\nabla_\gamma$$
$$- C_{\dot\alpha\dot\beta}[\, W_{\alpha\beta\gamma\delta} + i\tfrac{1}{4}C_{\gamma(\alpha|}(\nabla_{|\beta)}{}^{\dot\epsilon}H_{\delta\dot\epsilon}) + \tfrac{1}{24}C_{\gamma(\alpha}C_{\beta)\delta}([\nabla^\epsilon, \overline{\nabla}^{\dot\epsilon}]H_{\epsilon\dot\epsilon})\,]\mathcal{M}^{\gamma\delta}$$
$$+ \tfrac{1}{4}C_{\alpha\beta}[\,([\nabla_\gamma, \overline{\nabla}_{(\dot\alpha|}]H_{\delta|\dot\beta)}) + i(\nabla_{\gamma(\dot\alpha|}H_{\delta|\dot\beta)})\,]\mathcal{M}^{\gamma\delta}\} + \text{h.c.}\,\Big] . \tag{2.2d}$$

This formulation [8] was first presented in 1988 [7], and was motivated by a study of the superspace geometry associated with a set of vertex operators [14] constructed using superconformal field methods. On this basis, Cecotti *et al.* [14][6] concluded that the pure off-shell supergravity limit of 4D, $N = 1$ string theory was the new minimal theory. We note that the $\beta$FFC system is equivalent to the earlier but different constraint set for off-shell 4D, $N = 1$ superspace derived from the heterotic string given in ref.[13].

---

[5] Our (anti)symmetrizations in this section of superspace is the same as in [15], *e.g.*, $U_{[A}V_{B)} \equiv U_A V_B \pm U_B V_A$.

[6] Actually, Cecotti *et al.* only used component fields discussions and the equations above were found to be equivalent to their results.

To see the physical content more explicitly, we identify the $\theta = 0$ components as $A_a(z)| = A_a(x)$, $H_a(z)| = H_a(x)$. Accordingly, the $U_A(1)$ superfield strength $F_{MN} \equiv \partial_{[M}A_{N)}$ satisfies the usual $F$-Bianchi identity

$$\nabla_{[A}F_{BC)} - T_{[AB|}{}^D F_{D|C)} \equiv 0 \ . \tag{2.3}$$

The explicit form of $F_{BC}$ can be read from (2.3) from the coefficients of the $\mathcal{Y}$ generator,

$$F_{\alpha\dot\alpha} = iH_{\alpha\dot\alpha} \ , \quad F_{\alpha,\beta\dot\beta} = -\nabla_\beta H_{\alpha\dot\beta} \ , \quad F_{\alpha\dot\alpha,\beta\dot\beta} = \left[ \tfrac{1}{2}C_{\alpha\beta}\nabla^\gamma \overline\nabla_{(\dot\alpha|}H_{\gamma|\dot\beta)} + \text{h.c.} \right] . \tag{2.4}$$

The third-rank antisymmetric superfield strength defined by $H_{MNP} \equiv (1/2)\partial_{[M}a_{NP)}$, with the relation $H_{abc} \equiv \epsilon_{abc}{}^d H_d$, consistently satisfies the $H$-Bianchi identity

$$\tfrac{1}{6}\nabla_{[A}H_{BCD)} - \tfrac{1}{4}T_{[AB|}{}^E H_{E|CD)} \equiv 0 \ , \tag{2.5}$$

where the constraints for $H_{ABC}$ are given by

$$H_{\alpha\beta\gamma} = H_{\alpha\beta\dot\gamma} = H_{\alpha\dot\beta\gamma} = H_{\alpha\beta c} = H_{\alpha\dot\beta c} - i\tfrac{1}{2}C_{\alpha\gamma}C_{\dot\beta\dot\gamma} = 0 \ , \tag{2.6a}$$

$$H_{\alpha bc} = 0 \ , \quad H_{abc} = i\tfrac{1}{4}\left[ C_{\beta\gamma}C_{\dot\alpha(\dot\beta|}H_{\alpha|\dot\gamma)} - C_{\dot\beta\dot\gamma}C_{\alpha(\beta}H_{\gamma)\dot\alpha} \right] \ , \tag{2.6b}$$

$$T_{ab}{}^c = H_{ab}{}^c = \epsilon_{ab}{}^{cd}H_d \ . \tag{2.6c}$$

Using the commutator algebra as well as the superfield strengths, we obtain the supersymmetry transformation rule for these components as [15]

$$\delta_Q A_a = -i\tfrac{1}{2}\epsilon^\alpha(\sigma^b)_\alpha{}^{\dot\beta}(\sigma_a)^\gamma{}_{\dot\beta}\mathcal{R}_{b\gamma} + i\tfrac{1}{2}\overline\epsilon^{\dot\alpha}(\sigma^b)^\beta{}_{\dot\alpha}(\sigma_a)_\beta{}^{\dot\gamma}\overline{\mathcal{R}}_{b\dot\gamma} \ , \tag{2.7a}$$

$$\delta_Q H_a = i\epsilon^\alpha \mathcal{R}_{a\alpha} - i\overline\epsilon^{\dot\alpha}\overline{\mathcal{R}}_{a\dot\alpha} \equiv i(\epsilon\mathcal{R}_a) - i(\overline\epsilon\overline{\mathcal{R}}_a) \ , \tag{2.7b}$$

$$\delta_Q a_{ab} = i(\epsilon\sigma_{[a}\overline\psi_{b]}) + i(\overline\epsilon\overline\sigma_{[a}\psi_{b]}) \ . \tag{2.7c}$$

where we are using the notation in which the $\sigma$-matrices are manifest, e.g., $V_{\alpha\dot\alpha} \equiv V_a(\sigma^a)_{\alpha\dot\alpha}$, that is also consistent with the notations in [15], in order to make the comparison with new minimal supergravity [2] more straightforward. The quantity $\mathcal{R}_a$ is the l.h.s. of the gravitino field equation defined by the gravitino field strength:

$$\mathcal{R}_{a\alpha} \equiv -\tfrac{1}{2}\epsilon_a{}^{bcd}(\sigma_b)_\alpha{}^{\dot\beta}\overline{\mathcal{R}}_{cd\dot\beta} \ , \quad \overline{\mathcal{R}}_{a\dot\alpha} \equiv \tfrac{1}{2}\epsilon_a{}^{bcd}(\sigma_b)^\beta{}_{\dot\alpha}\mathcal{R}_{cd\beta} \ , \tag{2.8}$$

and $\mathcal{R}_{ab}{}^{\underline\gamma} \equiv T_{ab}{}^{\underline\gamma}|$ is the gravitino field strength. Here we use the *underlined* spinorial indices, in order to include the both chiralities: $\underline\alpha = (\alpha,\dot\alpha)$. Combined with the ordinary component transformation rule for the vierbein $\delta_Q e_a{}^m$ and gravitino $\delta_Q \psi_a{}^{\underline\beta}$ [15], eq. (2.7) implies that the above system (2.2) is identified with the new minimal supergravity theory with the field content $(e_m{}^a, \psi_m{}^{\underline\alpha}, a_{mn}, A_m)$ [2]. The only minor difference is that the $U_A(1)$-gauge field in [3] is the sum of $A_m + H_m$ instead of just $A_m$.

Since this formulation (or any formulation of pure 4D, $N = 1$ supergravity) of supergravity [7][8] does *not* contain a propagating dilaton, 2-form axion or dilatino, this pure supergravity sector must be coupled further to another multiplet with the component fields $(L, G_a, \chi)$ (a linear multiplet of the usual type) in order to accommodate the expected fields arising from superstring theory. From these results above, we conclude that the $\beta$FFC system of $N = 1$ supergravity in 4D described in [7][8] corresponds to new minimal supergravity [2] coupled to a compensating chiral multiplet (CCM).

## 3. Kinetic Lagrangian for CCM

We wish to probe $\beta$FFC supergravity as new minimal supergravity coupled to a CCM. The importance of the compensating chiral superfield has been recognized since it first appeared in the published literature [16], and more recently its role as an order parameter for supersymmetry-breaking has come to the fore [17]. Many times in the past, the CCM was often regarded as simply a curiosity of the 'formalism' of superfield supergravity [16]. It is now obvious that this 'formality' has powerful implications for phenomenology based on supersymmetrical extensions of the standard model.

We need to specify the system more explicitly to study the physical structure of the $\beta$FFC system. For this purpose, we start with the kinetic Lagrangian for a chiral multiplet coupled to $N = 1$ new minimal supergravity in 4D [2][3]. The field content of the chiral multiplet is the standard one $\Xi \equiv (A, B, \chi, F, G)^7$, *i.e.*, the scalar $A$, the pseudoscalar $B$, the Majorana spinor $\chi$, the scalar $F$ and pseudoscalar $G$ auxiliary fields. This multiplet is coupled to the multiplet of new minimal supergravity $(e_\mu{}^m, \psi_\mu, A_\mu, a_{\mu\nu})$ [2][3], where in addition to the vierbein $e_\mu{}^m$ and the gravitino $\psi_\mu$, there are auxiliary fields; a vector $A_\mu$ and an antisymmetric tensor 'auxiliary'[8] field $a_{\mu\nu}$, where the former also gauges the local $U_A(1)$ symmetry of the system [2][3].

We start with the kinetic Lagrangian for the CCM with the characteristic 'wrong' overall sign for its kinetic terms:[9]

$$\begin{aligned}
(-1)\, e^{-1} \mathcal{L}_{\text{CCM}} =\ & \tfrac{1}{4} e^{-1} q (A^2 + B^2) \mathcal{L}_{\text{SG}} - i\tfrac{1}{4} q \overline{\chi}(A - i\gamma_5 B)\gamma^\mu \widehat{\mathcal{R}}_\mu \\
& - \tfrac{1}{2}(\widehat{\mathcal{D}}_\mu A)^2 - \tfrac{1}{2}(\widehat{\mathcal{D}}_\mu B)^2 - i\tfrac{1}{2}(\overline{\chi}\gamma^\mu \widehat{\mathcal{D}}_\mu \chi) - \tfrac{1}{2} F^2 - \tfrac{1}{2} G^2 \\
& + \tfrac{3}{4}(\overline{\chi}\gamma_5 \gamma^\mu \chi)\widehat{V}_\mu + i\tfrac{1}{2} \overline{\chi}\gamma^\mu (F - i\gamma_5 G)\psi_\mu + \tfrac{1}{2}\overline{\chi}[\widehat{\slashed{\mathcal{D}}}(A - i\gamma_5 B)]\gamma^\mu \psi_\mu \\
& - \tfrac{1}{4} e^{-1} \epsilon^{\mu\nu\rho\sigma}(A\partial_\mu B - B\partial_\mu A + \tfrac{1}{2}\overline{\chi}\gamma_5 \gamma^\mu \chi)\, \partial_\nu a_{\rho\sigma} \quad.
\end{aligned} \quad (3.1)$$

We are mostly following the same notation as in [3], *except for* the $\gamma_5$-matrix which is now $(\gamma_5)^2 = +I$, the usage of $m, n, \cdots$ indices for local Lorentz frames, and the usage of the

---

[7]This supermultiplet was denoted by the symbol $\Lambda$ in [7]. See equation (4.4) there.

[8]We have put the quotation mark for 'auxiliary', because we will see that this field eventually becomes 'propagating' in our formulation of $\beta$FFC supergravity with its CCM.

[9]Note that our signature is $(+, -, -, -)$ as in [3], so that the 'physical' sign for a scalar kinetic term is positive: $+ (1/2)(\partial_\mu \varphi)^2$.

*hats* elucidating the supercovariantizations of derivatives, as the traditional convention. The number $q$ is the $U_A(1)$ charge of the chiral multiplet, which is nonzero $(q \neq 0)$ in our $\beta$FFC formulation. The field $\widehat{V}_m$ is the dual of the supercovariant field strength of $a_{\mu\nu}$, corresponding to $H_a$ in (2.7b), and $\widehat{\mathcal{R}}_\mu$ is the l.h.s. of the gravitino field equation [18] defined by

$$\widehat{V}^\mu \equiv e^{-1}\epsilon^{\mu\nu\rho\sigma}\left(\tfrac{1}{4}\partial_\nu a_{\rho\sigma} - i\tfrac{1}{2}\overline{\psi}_\nu\gamma_\rho\psi_\sigma\right) \equiv \tfrac{1}{4}e^{-1}\epsilon^{\mu\nu\rho\sigma}\widehat{\mathcal{D}}_\nu a_{\rho\sigma} \ , \tag{3.2a}$$

$$\widehat{\mathcal{R}}^\mu \equiv e^{-1}\epsilon^{\mu\nu\rho\sigma}\gamma_5\gamma_\nu \widehat{\mathcal{R}}_{\rho\sigma} \ , \tag{3.2b}$$

where $\widehat{R}_{\mu\nu} \equiv \widehat{\mathcal{D}}_\mu\psi_\nu - \widehat{\mathcal{D}}_\nu\psi_\mu$ is the supercovariant field strength of the gravitino [18]. The $U_A(1)$-covariant quantities $\mathcal{D}_\mu A$ and $\mathcal{D}_\mu B$ are defined by

$$\mathcal{D}_\mu A \equiv \partial_\mu A - qA_\mu B \ , \quad \mathcal{D}_\mu B \equiv \partial_\mu B + qA_\mu A \ . \tag{3.3}$$

The quantity $\mathcal{L}_{SG}$ in (3.1) is the Lagrangian of new minimal supergravity [2]

$$e^{-1}\mathcal{L}_{SG} = -\tfrac{1}{2}(\widehat{R} + 6\widehat{V}_\mu^2) + \tfrac{1}{2}(\overline{\psi}_\mu\gamma^\mu\gamma^\nu\widehat{\mathcal{R}}_\nu) - e^{-1}\epsilon^{\mu\nu\rho\sigma}A_\mu\partial_\nu a_{\rho\sigma} \ , \tag{3.4}$$

where $\widehat{R}$ is the fully supercovariantized scalar curvature [18][3]. This form of the action obscures the solely auxiliary and non-physical role of $a_{\rho\sigma}$. A re-definition of $A_\mu$ by an appropriate 'shift' eliminates the apparent kinetic terms of $a_{\rho\sigma}$ above.

The supersymmetry and the $U_A(1)$ transformations are dictated by the rule[10]

$$\begin{aligned}
\delta e_\mu{}^m &= -2i(\overline{\epsilon}\gamma^m\psi_\mu) \ , \\
\delta\psi_\mu &= \mathcal{D}_\mu\epsilon + i\gamma_5\epsilon\widehat{V}_\mu - \tfrac{1}{2}\gamma_5\gamma_{\mu\nu}\epsilon\widehat{V}^\nu - i\alpha\gamma_5\psi_\mu \ , \\
\delta A_\mu &= -i\tfrac{1}{4}(\overline{\epsilon}\gamma_5\gamma_\mu\gamma^\nu\widehat{\mathcal{R}}_\nu) + \partial_\mu\alpha \ , \\
\delta\widehat{V}_m &= i\tfrac{1}{2}(\overline{\epsilon}\gamma_5\widehat{\mathcal{R}}_m) \ ,
\end{aligned} \tag{3.5a}$$

$$\begin{aligned}
\delta A &= (\overline{\epsilon}\chi) + \alpha qB \ , \quad \delta B = -i(\overline{\epsilon}\gamma_5\chi) - \alpha qA \ , \\
\delta\chi &= -(F - i\gamma_5 G)\epsilon - i\widehat{\slashed{\mathcal{D}}}(A - i\gamma_5 B)\epsilon - i\alpha(q-1)\gamma_5\chi \ , \\
\delta F &= i(\overline{\epsilon}\widehat{\slashed{\mathcal{D}}}\chi) - \tfrac{1}{2}(\overline{\epsilon}\gamma_5\gamma^m\chi)\widehat{V}_m + i\tfrac{1}{4}q\overline{\epsilon}(A - i\gamma_5 B)\gamma^\mu\widehat{\mathcal{R}}_\mu + (2-q)\alpha G \ , \\
\delta G &= (\overline{\epsilon}\gamma_5\widehat{\slashed{\mathcal{D}}}\chi) + i\tfrac{1}{2}(\overline{\epsilon}\gamma^m\chi)\widehat{V}_m + \tfrac{1}{4}q\overline{\epsilon}\gamma_5(A - i\gamma_5 B)\gamma^\mu\widehat{\mathcal{R}}_\mu - (2-q)\alpha F \ ,
\end{aligned} \tag{3.5b}$$

where $\epsilon(x)$ and $\alpha(x)$ are respectively the parameters of supersymmetry and the $U_A(1)$ transformations.

It is convenient for later purposes to rewrite this system in terms of the complex fields

$$\exp\phi \equiv A + iB \ , \quad f \equiv e^{-(\phi+\phi^*)/2}(F - iG) \ . \tag{3.6}$$

---

[10]Readers have to be cautious about some difference in normalization of this Sohnius-Wess notation [2][3] from our superspace notation in the previous section, *e.g.*, $T_{\alpha\dot{\beta}}{}^c = +i(\sigma^c)_{\alpha\dot{\beta}}$, while from (3.5a) it is $T_{\alpha\dot{\beta}}{}^c = +2i(\sigma^c)_{\alpha\dot{\beta}}$, or similarly for the normalization in (2.8) compared with (3.2b).

For example, the $U_A(1)$ transformations for the new spinless fields are simplified as

$$\delta_\alpha \phi = -i\alpha q \, , \qquad \delta_\alpha f = +i\alpha(2-q)f \, , \qquad (3.7)$$

so that an $U_A(1)$ covariant derivative is defined by

$$\mathcal{D}_\mu \phi \equiv \partial_\mu \phi + iqA_\mu \, , \qquad (3.8)$$

where the last term *independent* of $\phi$ defines a 'minimal' coupling. These equations indicate that the pure imaginary part of $\phi$ is not really physical but can be gauged away by the $U_A(1)$ symmetry, as will be seen in the next section. The Lagrangian (3.1) is rewritten as

$$\begin{aligned}(-1)\,e^{-1}\mathcal{L}_{\text{CCM}} =\ & +\tfrac{1}{4}q\,e^{\phi+\phi^*}\left[-\tfrac{1}{2}\widehat{R} - 3\widehat{V}_\mu^2 + \tfrac{1}{2}(\overline{\psi}_\mu\gamma^\mu\gamma^\nu\widehat{\mathcal{R}}_\nu) - e^{-1}\epsilon^{\mu\nu\rho\sigma}A_\mu\partial_\nu a_{\rho\sigma} - 2q^{-1}|f|^2\right] \\ & -\tfrac{1}{2}e^{\phi+\phi^*}|\widehat{\mathcal{D}}_\mu\phi|^2 - i\tfrac{1}{2}(\overline{\chi}\gamma^\mu\widehat{\mathcal{D}}_\mu\chi) - i\tfrac{1}{4}qe^\phi(\overline{\chi}^-\gamma^\mu\widehat{\mathcal{R}}_\mu^+) - i\tfrac{1}{4}qe^{\phi^*}(\overline{\chi}^+\gamma^\mu\widehat{\mathcal{R}}_\mu^-) \\ & +\tfrac{3}{4}(\overline{\chi}\gamma_5\gamma^\mu\chi)V_\mu + i\tfrac{1}{2}e^{(\phi+\phi^*)/2}[f^*(\overline{\chi}^+\gamma^\mu\psi_\mu^-) + f(\overline{\chi}^-\gamma^\mu\psi_\mu^+)] \\ & +\tfrac{1}{8}e^{-1}\epsilon^{\mu\nu\rho\sigma}\Big[ie^{\phi+\phi^*}\{\partial_\mu(\phi-\phi^*)\} - (\overline{\chi}\gamma_5\gamma_\mu\chi)\Big]\partial_\nu a_{\rho\sigma} \\ & +\tfrac{1}{2}e^\phi\overline{\chi}^+(\widehat{\mathcal{D}}\phi)\gamma^\mu\psi_\mu^+ + \tfrac{1}{2}e^{\phi^*}\overline{\chi}^-(\widehat{\mathcal{D}}\phi^*)\gamma^\mu\psi_\mu^- \, , \end{aligned} \qquad (3.9)$$

while the transformation rule (3.5b) is now

$$\begin{aligned}\delta\phi &= 2e^{-\phi}(\overline{\epsilon}^+\chi^+) - i\alpha q \, , & \delta\phi^* &= 2e^{-\phi^*}(\overline{\epsilon}^-\chi^-) + i\alpha q \, , \\ \delta\chi &= -e^{(\phi+\phi^*)/2}(f\epsilon^+ + f^*\epsilon^-) - ie^\phi\gamma^\mu\epsilon^-\widehat{\mathcal{D}}_\mu\phi - ie^{\phi^*}\gamma^\mu\epsilon^+\widehat{\mathcal{D}}_\mu\phi^* - i\alpha(q-1)\gamma_5\chi \, , \\ \delta f &= e^{-(\phi+\phi^*)/2}\Big[-2i(\overline{\epsilon}^-\widehat{\mathcal{D}}\chi^+) + (\overline{\epsilon}^-\gamma^m\chi^+)\widehat{V}_m + i\tfrac{1}{2}qe^\phi(\overline{\epsilon}^-\gamma^\mu\widehat{\mathcal{R}}_\mu^+) + i\tfrac{1}{2}qe^{\phi^*}(\overline{\epsilon}^+\gamma^\mu\widehat{\mathcal{R}}_\mu^-)\Big] \\ & \quad - (e^{-\phi}\overline{\epsilon}^+\chi^+ + e^{-\phi^*}\overline{\epsilon}^-\chi^-)f + i\alpha(2-q)f \, . \end{aligned} \qquad (3.10)$$

In these expressions, the superscripts $\pm$ on spinorial fields denote the usual chiralities by the projections $(I \pm \gamma_5)/2$.

### 4. From New Minimal to 4D, $N = 1$ Minimal Stringy Supergravity

As mentioned above, the kinetic term for $\Xi$ has the wrong sign. Unless some unexpected phenomena occur, the system described thus far cannot be unitary as it must be plagued with ghosts. We first look into the role played by $A_\mu$, the auxiliary gauge field for the $U_A(1)$ symmetry, in order to see how the system of new minimal supergravity is modified by the chiral compensator. To this end, we study the purely bosonic part of (3.9):

$$\begin{aligned}(-1)e^{-1}\mathcal{L}_{\text{Bos}} &= e^{\phi+\phi^*}\Big\{-\tfrac{1}{8}qR + \tfrac{1}{32}q(H_{\rho\sigma\tau})^2 - \tfrac{1}{2}|\mathcal{D}_\mu\phi|^2 - \tfrac{1}{2}|f|^2 \\ &\quad -\tfrac{1}{12}qe^{-1}\epsilon^{\mu\nu\rho\sigma}\Big(A_\mu - i\tfrac{1}{2}q^{-1}[\partial_\mu(\phi-\phi^*)]\Big)H_{\rho\sigma\tau}\Big\} \\ &= e^{\phi+\phi^*}\Big\{-\tfrac{1}{8}qR + \tfrac{1}{96}(3q-2)(H_{\rho\sigma\tau})^2 - \tfrac{1}{8}[\partial_\mu(\phi+\phi^*)]^2 - \tfrac{1}{2}(q^2\widetilde{A}_\mu^2 + |f|^2)\Big\} \, , \qquad (4.1)\end{aligned}$$

where $H_{\rho\sigma\tau}$ and $\widetilde{A}_\mu$ are defined by

$$H_{\rho\sigma\tau} \equiv 3\,\partial_{[\rho} a_{\sigma\tau]} \quad , \quad \widetilde{A}_\mu \equiv A_\mu - i\tfrac{1}{2} q^{-1} \partial_\mu(\phi - \phi^*) + \tfrac{1}{4} q^{-1} e^{-1} \epsilon^{\mu\nu\rho\sigma} \partial_\nu a_{\rho\sigma} \quad . \quad (4.2)$$

Note the important fact that the 'mass term' $\approx (1/2) q^2 \widetilde{A}_\mu^2$ emerges at the bilinear order, due to the 'minimal' coupling in $\mathcal{D}_\mu \phi$ from (3.8). We note that $\phi + \phi^*$ appears in the exponent as the common factor, which can be identified with the dilaton field of superstring theories.

The kinetic term of the component $i(\phi - \phi^*)$ has completely disappeared! This is no wonder considering (3.7) that the pure imaginary part of $\phi$ can be gauged away by the $U_A(1)$ symmetry. It is now clear that the elimination of the auxiliary field $\widetilde{A}_\mu$ results in the generation of the kinetic term for $a_{\mu\nu}$, which replaces the original dynamical degree of freedom of the component $i(\phi - \phi^*) \approx B$ at the lowest order. To put it differently, this implies that the non-dynamical component $i(\phi - \phi^*)$ can be understood as a Nambu-Goldstone field, that is absorbed by the 2-form axion $a_{\mu\nu}$. As observed before [7], the $U_A(1)$ symmetry is induced by the well known $U(1)$ current in superconformal field theory that gives rise to 4D, $N = 1$ target space supersymmetry. Alternating, $B$ is the corresponding target space Nambu-Goldstone field of the symmetry generated by this current.

Interestingly, the coefficient of the kinetic term for $a_{\mu\nu}$ is $(1/96)(3q-2)$, so that unitarity is realized only if the $U_A(1)$ charge is bounded from below

$$q > \tfrac{2}{3} \quad . \quad (4.3)$$

In particular, the minimal integral value is $q = 1$. (This charge has been noted before in the works of [13] and [19] where it was denoted by the symbols $m$ and $n$, respectively.)

We next check the unitarity conditions for the kinetic term of the component $\phi + \phi^*$. This can be seen by an appropriate Weyl-rescaling from the Brans-Dicke Lagrangian: $-(1/8)\,q\,e\,e^{\phi+\phi^*} R$ to the standard Hilbert Lagrangian: $-(1/2)R$. This is performed by

$$e_\mu{}^m \longrightarrow \tfrac{2}{\sqrt{q}} e^{-(\phi+\phi^*)/2} e_\mu{}^m \quad , \quad (4.4)$$

yielding

$$e^{\phi+\phi^*}\left( -\tfrac{1}{8} q\,e\,R - \tfrac{1}{8} e [\partial_\mu(\phi+\phi^*)]^2 \right)$$
$$\longrightarrow -\tfrac{1}{2} eR + \left(\tfrac{3q-2}{4q}\right) e\,[\partial_\mu(\phi+\phi^*)]^2 + \text{(4-divergence)} \quad . \quad (4.5)$$

We re-encounter the condition (4.3) for the unitarity of the kinetic term of $\phi + \phi^* \approx L$.

Note also that the original auxiliary fields $F$ and $G$ in (4.1) are now regarded as $S$ and $P$ auxiliary fields with the same signature as in old minimal supergravity [1][18], while the negative signature of the $A_\mu^2$-term is also in agreement. From these developments, our resulting bosonic Lagrangian (4.1) has a spectrum that is identical to that of old minimal supergravity $(e_\mu{}^m, \psi_\mu, S, P, \widetilde{A}_\mu)$ coupled to the linear multiplet $(L, a_{\mu\nu}, \chi)$. In particular, the component $L \approx \phi + \phi^*$ is to be identified as the 'dilaton' with an overall exponential coupling to the total Lagrangian. Thus we see that our results comply also with the statement in [13][11] that the antisymmetric tensor field $a_{\mu\nu}$ is to be dynamical, so that the real background for superstrings and heterotic strings with $N=1$ target space supersymmetry is $\beta$FFC supergravity [7], but *not* new minimal [2] supergravity. The condition $q > 2/3$ is also consistent with unitarity, *e.g.*, the critical value $q = 2/3$ yields the disappearance of all the kinetic terms in the multiplet $(L, a_{\mu\nu}, \chi)$. In fact, after the Weyl-rescaling (4.4) together with the elimination of the mixture between $\psi_\mu$ and $\chi$ by

$$\psi_\mu \to \psi_\mu - i \frac{1}{\sqrt{q}} \gamma_\mu \chi \ , \tag{4.6}$$

we can easily confirm that the kinetic terms for $\psi_\mu$ and $\chi$ are

$$\tfrac{\sqrt{q}}{2} e e^{-(\phi+\phi^*)/2} (\overline{\psi}_\mu \mathcal{R}^\mu) \ + \ \tfrac{2}{\sqrt{q}} \left( \tfrac{3q-2}{q} \right) i e e^{-(\phi+\phi^*)/2} (\overline{\chi} \gamma^\mu \mathcal{D}_\mu \chi) \ , \tag{4.7}$$

Therefore, the kinetic term of $\chi$ also vanishes at the critical value $q = 2/3$. Note that our mechanism of one spin-0 scalar field (Nambu-Goldstone boson) eaten up by another spin-0 antisymmetric tensor field is essentially similar to an example of the Goldstone mechanism.

*We assert the action in (3.9)* <u>must</u> *be the component level formulation for 4D, N = 1 $\beta$FFC supergravity and corresponds to a term in the effective action of superstring/M-theory.*

## 5. Dilatonic Superpotential

We seem also to be able to do something with the $\beta$FFC formulation that has *never* to our knowledge been possible before. It is possible to introduce a dilaton superpotential[12] in our system, that may or may not break the $U_A(1)$ symmetry depending on the explicit form of the dilatonic superpotential. For example, this can be done by adding the mass term [3]

$$\mathcal{L}_{\mathrm{m}} \equiv -\tfrac{1}{2} m\, e |\Xi \cdot \Xi|_F \ , \tag{5.1}$$

for the CCM with a particular value of $q = 1$, with the dot for the product of chiral multiplets in local tensor calculus in new minimal supergravity [3], or we can add the cubic potential term [3]

$$\mathcal{L}_{\Xi^3} \equiv -\tfrac{1}{3} g e |\Xi \cdot \Xi \cdot \Xi|_F \ , \tag{5.2}$$

for another particular value of $q = 2/3$. (However, as we have seen there are unitarity problems for this choice of the $U_A(1)$ charge.) In fact, a more general dilatonic superpotential of the form

$$\mathcal{L}_\mathcal{N} \equiv e |\mathcal{N}(\Xi)|_F \tag{5.3}$$

---

[11] There has been some controversy about the 'true' off-shell version of supergravity derived from $N=1$ strings in 4D [13][14]. We believe that the result of our present paper should provide clarification of the issue by giving the explicit link between new minimal [2], old minimal [1] and $\beta$FFC [7] supergravity.

[12] This feature is outside of any conventional discussion of superstrings.

may be added and this, except for the special choices, explicitly breaks the $U_A(1)$ symmetry. We emphasize that the chiral compensator $\Xi$, unlike the traditional chiral compensator introduced in [16], *only* compensates for the local $U_A(1)$ symmetry. It is this feature that allows the general dilatonic superpotential in (5.3) to be introduced.

## 6. Concluding Remarks

We have given the first component level presentation of the link between the old minimal [1], new minimal [2][3] and $\beta$FFC formulations [7] of off-shell 4D, $N = 1$ supergravity. Starting with a kinetic Lagrangian for a CCM and coupling to new minimal supergravity requires a number of subtle but substantial redefinitions that lead to a unitary spectrum. For example, the kinetic term with the 'wrong' sign for the scalar field $A$, will not pose any problem, due to the Weyl-rescaling to be performed to adjust the Brans-Dicke term into the standard Einstein-Hilbert Lagrangian. The elimination of the auxiliary gauge field $A_\mu$ for the axial $U_A(1)$ symmetry results in the absorption of the kinetic term of the pseudoscalar $B$, while generating a new kinetic term for the antisymmetric tensor $a_{\mu\nu}$, causing it to become propagating. Interestingly, we have found that unitarity in the $\beta$FFC formulation is achieved only with the restriction[13] $q > 2/3$ for the $U_A(1)$ charge $q$ of the chiral compensating multiplet $\Xi$. We again emphasize that the 'wrong' sign originates from 4D, $N = 1$ BRST superstring arguments [13], *i.e.,* the chiral compensator has its origins in the ghost number zero sector of the superstring $b$-$c$ ghosts-anti-ghost system.

We also note that the introduction of some mechanism for $U_A(1)$ symmetry-breaking, such as the introduction of the CCM can also be seen from another argument. Shortly after the introduction of the new minimal theory, it was shown that new minimal supergravity necessarily contains an 'auxiliary field anomaly' [20] and thus is *not* a quantum mechanically consistent theory. So it is actually compulsory that the $U_A(1)$-symmetry in new minimal supergravity be broken in order to arrive at a consistent theory. The CCM implements this breaking.

We also emphasize that the $\beta$FFC formulation is the first 'minimal' system, in which there is a dilaton that can accommodate a superpotential and *simultaneously* the antisymmetric tensor (axion) is present. These two fields are essential for superstring theory, because the antisymmetric tensor is indispensable for the Green-Schwarz anomaly cancellation mechanism [21], not to mention the importance of the dilaton in superstring theory [9]. For example, in the old fashioned way, we couple a chiral multiplet to old minimal supergravity with the dilaton in the chiral multiplet. But this old formulation needs an additional linear multiplet to accommodate the antisymmetric tensor for Green-Schwarz mechanism. On the other hand, in what is called the 'linear formulation', a linear multiplet with a propagating antisymmetric tensor is coupled to old minimal supergravity. But it is not known how to introduce a dilaton superpotential, without breaking supersymmetry, in such a formulation either. From these considerations, we see that the $\beta$FFC formulation is the first formulation that has the following features, *i.e.*, (i) manifest local supersymmetry, (ii) a 2-form for implementing the Green-Schwarz mechanism, (iii) a supersymmetry-breaking order parameter chiral superfield, and (iv) the possibility to admit a dilatonic superpotential.

---

[13]Although it is beyond the scope of this work, we cannot but wonder about the 'stringy' origin of the condition $q > 2/3$.

We believe that our present paper provides a very beautiful resolution to the old puzzle [13][14] about the choice between old minimal [1] and new minimal [2] supergravity as the rigorously correct background of the 4D, $N=1$ superstring [13], and may open a new avenue for superstring/M-theory and supergravity.

> *"Now this is not the end. It is not even the beginning of the end. But it is, perhaps, the end of the beginning."* – Sir Winston L.S. Churchill.

**Acknowledgment:**

We thank W. Siegel for reminding us of the work in reference [13].

**Added Note in Proof**

After the completion of our work, it was brought to our attention that the dilatonic superpotential has previous appeared in the work of [22]. Additionally, it was shown there that the charge $q=1$ is chosen by superstring theory. This in turn implies that the dilatonic superpotential can only take the form in (5.1).